\documentclass[twocolumn,showpacs,preprintnumbers,amsmath,amssymb,superscriptaddress]{revtex4-2}
\usepackage{epsfig,bm}
\usepackage{ulem}
\usepackage{amsmath}
\usepackage{color}
\begin{document}
\title{Possible halo structure of $^{62,72}$Ca by
  forbidden-state-free locally peaked Gaussians}
\author{W. Horiuchi\email{whoriuchi@nucl.sci.hokudai.ac.jp}}
\affiliation{Department of Physics, Hokkaido University, Sapporo 060-0810, Japan}
\author{Y. Suzuki}
\affiliation{Department of Physics, Niigata University, Niigata 950-2181, Japan}
\affiliation{RIKEN Nishina Center, Wako 351-0198, Japan}
\author{M.~A. Shalchi}
\affiliation{Instituto de F\'isica Te\'orica, Universidade Estadual Paulista, 01140-070, S\~ao Paulo, SP, Brasil}
\author{Lauro Tomio}
\affiliation{Instituto de F\'isica Te\'orica, Universidade Estadual Paulista, 01140-070, S\~ao Paulo, SP, Brasil}

\begin{abstract}
  In order to efficiently describe nucleon orbits around a heavy core nucleus, 
  we propose locally peaked Gaussians
  orthogonalized to the occupied bound states in the core. 
  We show the advantage of those functions
  in both numerical stability and fast convergence 
  by taking examples of touchstone calcium isotopes
  $^{62,72}$Ca in $^{60,70}{\rm Ca}+n+n$ three-body models. Both weakly bound configurations and 
  continuum coupling effect are taken into account. 
  We evaluate  the neutron radii and the occupation probabilities of two-neutron configurations not only 
  for the ground state but also for some particle-bound excited states 
  by varying the strength of the core-neutron interaction. The emergence of 
  the halo structure in the ground state depends on the
  energy difference between $2s_{1/2}$  and $0g_{9/2}$ orbits. 
Two-neutron [consisting of $(s_{1/2})^2$ configuration] and one-neutron
[consisting of $(g_{9/2}s_{1/2})$ configuration] halo structures
of $^{62}$Ca can  
  coexist in narrow energy spacing provided that both of $2s_{1/2}$  
  and $0g_{9/2}$ orbits are almost degenerate and barely bound.
  The ground-state structure of $^{72}$Ca is likely to be a two-neutron halo,
  although its emergence depends on the position of the $2s_{1/2}$
  level.
\end{abstract}

\maketitle

\section{Introduction}

The landscape of exotic short-lived nuclei far from the stability line
extends to the neutron dripline up 
to Ne isotopes~\cite{Ahn19}.  In such extreme neutron excess, 
weakly bound neutrons
emerge and often lead to neutron-halo structure.
Exploring the halo nuclei has continued since the first discovery of 
the two-neutron halo nucleus $^{11}$Li~\cite{Tanihata85}. 
Other typical examples
of two-neutron halo nuclei include $^{6}$He~\cite{Tanihata85},  
$^{14}$Be~\cite{Tanihata88}, $^{19}$B~\cite{Suzuki99}, and
$^{22}$C~\cite{Horiuchi06,Tanaka10,Togano16}.
See also Ref.~\cite{Tanihata13} 
for more experimental and theoretical references on halo-nuclei studies.
Recently, by measuring its interaction cross section, $^{29}$F 
has been identified as the heaviest two-neutron halo nucleus~\cite{Bagchi20}.

The two-neutron halo structure has often been described using
a core plus two-neutron model.  When the core is 
a light nucleus, the two-neutron motion can be accurately described, e.g., as in $^{22}$C~\cite{Horiuchi06}. 
Within the search for evidence of Efimov states~\cite{1970Efimov}
in two-neutron $(2n)$ halo nuclei close 
to the neutron-core ($n$-core) unitary limit~\cite{1994Fedorov,2012Frederico}, 
the emergence of universal properties described by two-body observables and a three-body scaling parameter became quite clear
in the case of two neutrons bound to light-core nuclei.

Motivated by the recent measurement reported in Ref.~\cite{Bagchi20},
the dripline and near-dripline F
isotopes have been studied by three-body models:
For $^{29}$F, the 
$p$-wave two-neutron halo character of $^{27}{\rm F}+n+n$ was verified 
in Refs.~\cite{Singh20,Fortunato20,Casal20},
and a competition of halo and antihalo configurations was proposed 
in Ref.~\cite{Masui20} for $^{31}$F studied as $^{29}{\rm F}+n+n$.
By using the Gamow shell model these F isotopes were also studied 
in Ref.~\cite{2020Michel}.
Studying the existence of heavier neutron-halo nuclei is interesting as it will 
help to expose the  binding mechanism when increasing the mass number~\cite{Hagen13,Hove18}, 
which opens the possibility of verifying universal aspects expected to emerge in low-energy quantum 
systems, such as  ultra-cold atom-molecule heteronuclear systems~\cite{2012Frederico,2018Incao,2018Shalchi}.
However, a microscopic description becomes tougher 
as the core nucleus becomes heavier 
because the number of the occupied orbits in the core increases. 
Therefore, a reliable and efficient method is needed to describe
the two-neutron motion around the heavy core nucleus.
The complication of such calculation
is mainly due to the condition that requires 
the configuration space available to the valence neutrons 
to be orthogonal to all the  bound states occupied in the core
(orthogonality condition model~\cite{OCM1,OCM3}). 
To eliminate such bound states (called forbidden states) 
from a three-body solution, a pseudo potential method~\cite{Kukulin78}
has often been applied.
With increasing the mass number of the core nucleus, the number of the occupied orbits increases and the 
description of the valence neutron orbits becomes more complicated and often numerically unstable.

To describe the nucleon motion around the core, one can use  explicitly
correlated Gaussians~\cite{Varga95,SVM, Mitroy13}.
Due to the complexity of removing the forbidden states, however, 
its application is limited only to light nuclei. See, e.g.,
Refs.~\cite{Suzuki08, Horiuchi14}.
Obviously, a single-particle (sp) basis
is advantageous to eliminate the forbidden state. 
The application of Gaussian sp basis functions 
to the alpha decay of $^{212}$Po in the $^{208}{\rm Pb}+n+n+p+p$
model was in fact successfully made in Ref.~\cite{Varga94}. 
It turned out, however, that the description was not perfect
for the localized nucleon orbits, i.e., the 
alpha clustering near the nuclear surface. This 
is probably because the ordinary Gaussians, $r^l\exp(-a r^2)$, used there  
are not always enough to obtain large amplitude near the nuclear surface.  
In Refs.~\cite{Suzuki17a,Suzuki20}, locally peaked Gaussians (LPGs),  
$r^{2k+l}\exp(-a r^2)$, are proposed to describe the localized configurations.
By the additional $r^{2k}$ factor, the LPG basis allows one to describe not only 
damped short-ranged behavior~\cite{Varga95,SVM} but also large amplitude far from the center,
while keeping the advantage that the matrix elements of the LPG bases   
can easily be evaluated.

In this paper, we introduce a forbidden-state-free LPG (FFLPG) 
to efficiently describe neutron orbits around a heavy core nucleus.
This opens up the perspectives to study
more complicated multinucleon systems  around the heavy core.
To show its effectiveness, 
we apply the method to describe $^{62,72}$Ca in which
the emergence of the two-neutron halo 
was discussed~\cite{Hagen13,Hove18}.
The purpose of this paper is twofold:
(1) to establish an efficient way to describe both the short-ranged nodal 
behavior and enhanced amplitude beyond the nuclear surface, which is 
expected to occur in nucleon orbits around the heavy core, and
(2) to clarify the 
 conditions under which halo structure can emerge
 not only in the ground state
but also in some excited states of the 
$^{60,70}{\rm Ca}+n+n$ three-body systems.

As mentioned above, we focus on the neutron-rich Ca isotopes having
large number $N$ of neutrons. Although 
the halo structure is expected to appear for $N>40$, 
its structure or even its existence is under debate. 
Very little experimental information is available for the Ca isotopes: 
The heaviest, $^{60}$Ca, was confirmed in Ref.~\cite{Tarasov18}
but no information  other than its existence is available. 
Mass measurements have been done up to $^{57}$Ca~\cite{Michimasa18},
and the charge radii~\cite{Ruiz16} have been determined up to $^{52}$Ca,
and recently interaction cross sections have been measured up
to $^{51}$Ca~\cite{Tanaka20}.
It was conjectured by the coupled-cluster calculation that the neutron dripline 
of Ca isotopes is around $^{60}$Ca~\cite{Hagen12} and $^{62}$Ca has 
two-neutron halo structure with dominant $s$-wave~\cite{Hagen13}.
On the other hand 
energy-density-functional~\cite{Erler12,Forssen13},
shell-model~\cite{Coraggio20}, and in-medium similarity renormalization
group~\cite{Stroberg21} calculations predicted
that the dripline is around $^{70}$Ca.
The halo structure of $^{72}$Ca
was also predicted by the relativistic mean-field model~\cite{Meng02}.
The two-neutron halo structure in the ground state of $^{72}$Ca
and its relationship to the Efimov physics was studied
by a $^{70}{\rm Ca}+n+n$ model~\cite{Hove18}.
The dripline of Ca isotopes was predicted to be $^{72}$Ca based on 
a Bayesian analysis of the density functional theory
results~\cite{Neufcourt19}. Since there is still some ambiguity 
in determining the dripline, we leave its question open and study possible neutron-halo structure 
in the spectrum of both cases, $^{62,72}$Ca.

The paper is organized as follows.
Section~\ref{formulation.sec} presents
the Hamiltonian and some definitions needed to introduce
the present approach
and explains how to construct the FFLPG.
Section~\ref{results.sec} discusses our results.
First in Sec.~\ref{pot.sec}, the ${\rm Ca}+n$ potential
employed in this paper is investigated.
Section~\ref{tests.sec} is devoted
to test the power of this approach
to the $^{60}{\rm Ca}+n+n$ system.
Comparison with a standard projection method is presented.
Section~\ref{62Ca.sec} discusses the emergence
of various halo structures in the spectrum of $^{62}$Ca
based on the three-body results.
The case of $^{72}$Ca is presented 
in Sec.~\ref{72Ca.sec}.
Conclusions and future perspectives are given in
Sec.~\ref{conclusions.sec}

\section{Formulation}
\label{formulation.sec}

\subsection{Hamiltonian and variational calculation}

The Hamiltonian of a core$+2n$ system consists of the 
 $n$-core kinetic ($T$) and potential ($U$)
and the $n$-$n$ potential ($v$) terms,
\begin{align}
  H=\sum_{i=1}^2(T_i+U_i)+\frac{1}{Am_N}\bm p_1\cdot \bm p_2+v_{12},
\label{hamiltonian}
\end{align}
where $m_N$ and $A$ are the nucleon 
mass and the mass number of the core nucleus, respectively. 
We follow the cluster-orbital shell model approach~\cite{Suzuki88}, 
by reducing the three-body 
problem to a two-body problem using two independent 
$n$-core relative distance vectors, 
$\bm r_1$ and $\bm r_2$ 
with their respective momentum conjugates $\bm p_1$ and $\bm p_2$.
The center-of-mass kinetic energy is subtracted and
the corresponding kinetic energies are given by $T_{i}= {\bm p_i^2}/{(2\mu)}$
with $\mu= m_NA/(A+1)$.
Our choices for the $n$-core and $n$-$n$ potentials will be given later.

The total wave function with
the angular momentum $J$ and its $z$ component $M$
is expanded in terms of $K$ basis functions $\Phi_{JM}(\alpha_i)$ $(i=1,\dots, K)$:
\begin{align}
  \Psi_{JM}=\sum_{i=1}^{K} c_i\Phi_{JM}(\alpha_i),
\label{expand.eq}
\end{align}
where $\alpha_i$ denotes a set of variational parameters
for the $i$th basis. These basis functions
are not restricted to be orthogonal.
A set of linear coefficients $\bm{c}=(c_1, \dots, c_K)^t$
is determined variationally by solving the generalized eigenvalue problem
\begin{align}
  \mathcal{H}\bm{c}=E\mathcal{B}\bm{c},
\end{align}
where $\mathcal{H}$ and $\mathcal{B}$ are the Hamiltonian and overlap matrices
with elements defined by
\begin{align}
  \mathcal{H}_{ij}=\left<\Phi(\alpha_i)|H|\Phi(\alpha_j)\right>,
  \label{meham.eq}
\end{align}
and
\begin{align}
  \mathcal{B}_{ij}=\left<\Phi(\alpha_i)|\Phi(\alpha_j)\right>.
\label{meovl.eq}
\end{align}
An optimal set of $\alpha_i$ is selected
by the stochastic variational method (SVM)~\cite{Varga95,SVM}.
Its efficiency  was demonstrated by
a number of examples.
See, e.g., Refs.~\cite{Suzuki08,Mitroy13,Suzuki17b}.
We increase the basis one by one by selecting
the one that gives the lowest energy  among randomly generated
candidates until the energy convergence is met.
This procedure greatly reduces the total number of basis $K$, which  
 helps  the description of nucleon systems
around a heavy core, where the number of possible configurations is quite large.

\subsection{Definition of FFLPG}

The efficiency of a variational calculation strongly depends on a choice of basis functions.
Some basic requirements for a ``good'' basis include the following in the present case:  
(i) Weakly bound neutron orbits should be well described, because the system 
 may have large amplitude at and beyond the surface of the core nucleus. 
 (ii) The evaluation of the matrix elements in Eqs.~(\ref{meham.eq})
 and (\ref{meovl.eq})
 should be easy and extendable 
to systems with more neutrons bound to the core.
(iii) The removal of many forbidden states should be performed
numerically easily and stably.

The sp wave function with total (orbital) angular momentum $j$ ($l$),
 with their corresponding angular and spin wave functions,
$Y_{l}(\hat{\bm{r}})$ and $\chi_{1/2}$, is defined by
\begin{align}
\phi_{kljm}^{a}=\phi_{kl}^{a}(r)\left[Y_{l}(\hat{\bm{r}})\chi_{1/2}\right]_{jm},
\end{align}
where the LPG function is given by
\begin{align}
  \phi_{kl}^a(r)=N_{kl}\left(\frac{a^3}{\pi}\right)^{\frac{1}{4}}(\sqrt{a}\,r)^{2k+l}
  \exp\left(-\frac{1}{2}ar^2\right),
\label{LPG.eq}
\end{align}
Here, $N_{kl}$ is  the normalization constant
\begin{align}
  N_{kl}=\sqrt\frac{2^{2k+l+2}}{(4k+2l+1)!!},
\end{align}
and $a$ is a parameter related to the width of the LPG function.
We assume $k$ to be a non-negative integer for the sake of simplicity.  
The LPG with $k=0$ is nothing but the ordinary Gaussians.
Most of basic matrix elements between the LPG bases can be obtained 
analytically~\cite{Suzuki17a}. 
Since the LPG reaches a maximum at $r=\sqrt{(2k+l)/a}$, a suitable combination 
of $a$ and $k$ 
can describe such wave packets that are centered at different positions and have 
different widths. Because of this flexibility, the LPG can describe 
even linear chain structure~\cite{Suzuki17a}. Obviously the tail 
of a weakly bound orbit can be described well 
by a superposition of the LPG. This flexibility is vital 
to describe the sp orbits around a heavy core nucleus.

The LPG basis satisfies the requirements (i) and (ii).
Concerning the requirement (iii), we introduce a projection operator
onto the forbidden ($F$) space,
\begin{align}
  P_F=\sum_{n^\prime l^\prime  j^\prime \in {F}}\sum_{m'=-j'}^{j'}
  \left|\psi_{n^\prime l^\prime  j^\prime m'}\right> \left<\psi_{n^\prime l^\prime  j^\prime m'}\right|,
\end{align}
where the sum extends over all the orbits occupied in the core nucleus.
The forbidden states $\psi_{nl j m}$ are defined by the bound-state solutions of the one-body 
Hamiltonian, $T+U$. As will be seen later, they are very well 
approximated by the harmonic-oscillator (HO) functions 
\begin{align}
\psi_{nljm}^\nu=\psi_{nl}^\nu(r)\left[Y_{l}(\hat{\bm{r}})\chi_{1/2}\right]_{jm}
\end{align}
with the principal quantum number ($n$) and an appropriately chosen
oscillator parameter $\nu$. We 
use this approximation in what follows. 
The FFLPG sp orbit is defined by
\begin{align}
  \bar{\phi}_{kljm}^{a}
  &=(1-P_F)\phi_{kljm}^{a}\notag\\
  &=\phi_{kljm}^{a}-\sum_{n^\prime ; n^\prime lj\in F}
  \left<\psi_{n^\prime l}^\nu\right.\left|\phi_{kl}^{a}\right>
  \psi_{n^\prime ljm}^\nu\notag \\
 & \equiv  \bar{\phi}_{kl}^{a}(r) \left[Y_{l}(\hat{\bm{r}})\chi_{1/2}\right]_{jm}.
\label{FFLPG.eq}
\end{align}
Because  $\psi_{n'l}^{\nu}(r)$ is a combination of LPG's, $\phi_{k'l}^{\nu}(r) \ (k'=0,\ldots,n')$, $\left<\psi_{n^\prime l}^\nu\right.\left|\phi_{kl}^{a}\right>$ is readily obtained by using
\begin{align}
\left<\left.\phi_{k'l}^{a'}\right|\phi_{kl}^{a}\right>&=\frac{(2k+2k'+2l+1)!!}{\sqrt{(4k+2l+1)!!(4k'+2l+1)!!}}\notag \\
&\times \frac{\sqrt{a}^{2k+l+\frac{3}{2}}\sqrt{a'}^{2k'+l+\frac{3}{2}}}{(\frac{a+a'}{2})^{k+k'+l+\frac{3}{2}}},
\end{align}
which leads 
to easy determination of $\bar{\phi}_{kljm}^{a}$.

With the use of the sp basis defined above, we construct an antisymmetrized two-neutron basis
\begin{align}
  &\Phi_{JM}(a_1k_1l_1j_1;a_2k_2l_2j_2)\notag\\
  &=\frac{1}{\sqrt{2}}(1-P_{12})
  \left\{\left[\bar{\phi}_{k_1l_1j_1}^{a_1}(1) \bar{\phi}_{k_2l_2j_2}^{a_2}(2)
    \right]_{JM}\right\},
\label{2n.conf}
\end{align}
where $P_{12}$ exchanges the neutron labels 1 and 2,
and $[j_1 j_2]_{JM}$ denotes the tensor product.
Here, the set of variational parameters of the $i$th basis $\alpha_i$
in Eq.~(\ref{expand.eq}) stands for 
$\alpha_i=(a_{1,i}k_{1,i}l_{1,i}j_{1,i},a_{2,i}k_{2,i}l_{2,i}j_{2,i})$.
This two-neutron basis can easily be extended
to core$+$few nucleon systems by successively 
coupling  another nucleon one by one.

Note that the variational parameter $\alpha_i$ comprises eight variables: two 
continuous ones ($a_{1,i}$ and $a_{2,i}$) and six discrete ones. 
As will be shown later, both of short- and long-ranged LPG's have to 
be superposed to properly describe the asymptotics of the 
FF halo wave functions and also continuum states 
with high angular momenta ($l_{1,i}$ and $l_{2,i}$) have to be 
included to reach convergence. Therefore, discretizing each component of 
$\alpha_i$ on certain grids would lead to enormous basis dimension even for 
the present three-body system.  More crucial is that one has to 
take into account the possibility of producing two bound states 
with the same spin and parity. A way to overcome these difficult 
problems is to reduce the basis dimension by the SVM. An interested reader 
should refer to Chapter 4 of Ref.~\cite{SVM}. 

The above basis of Eq.~(\ref{2n.conf}) is completely FF. 
In contrast to this approach, a popular way of eliminating the forbidden components is 
to add a pseudo-potential to the Hamiltonian as in Ref.~\cite{Kukulin78},
\begin{align}
H\to H+\lambda \sum_{i=1}^2P_{F}(i),
\label{proj.eq}
\end{align}
and to attempt at reaching a stable eigenvalue 
by taking $\lambda\to\infty$. In practice, 
$\lambda \approx 10^4$ MeV is taken~\cite{Suzuki08}. 
This pseudo-potential method has the advantage of its simplicity.
We compare both approaches in the next section. 

\section{Application to ${\rm Ca}+n+n$ systems}
\label{results.sec}

\subsection{Choice of $n$-$^{60}$Ca potential}
\label{pot.sec}

We take the Woods-Saxon (WS) form for the $n$-$^{60}$Ca potential
\begin{align}
  U(r)=-V_0f_{\rm WS}(r)+V_1r_0^2\frac{1}{r}\frac{df_{\rm WS}(r)}{dr}(\bm{\ell}\cdot\bm{s}),
\label{n-60Ca.pot}
\end{align}
where $f_{\rm WS}(r)=\left[1+\exp\left(r-R_{\rm WS}\right)/a_{\rm WS}\right]^{-1}$
with $R_{\rm WS}=r_0A^{1/3}$ and $r_0=1.27$ fm.
We use two sets for the diffuseness 
parameter: One is a standard one, $a_{\rm WS}=0.67$ fm (set A), and the other  
is a larger one, $a_{\rm WS}=0.80$ fm (set B). 

The $^{60}$Ca core nucleus is assumed to have the $N=40$ closed configuration, 
that is, the occupied orbits include $0s_{1/2}$, $0p_{3/2}$, $0p_{1/2}$,
$1s_{1/2}$, $0d_{5/2}$, $0d_{3/2}$, $1p_{3/2}$, $1p_{1/2}$, $0f_{7/2}$, $0f_{5/2}$.   
We generate these forbidden states by assuming the WS potential 
parameters of Eq.~(2-182) in Ref.~\cite{BM}.
Those potential parameters are $V_0=40.00$ MeV and $V_1=17.60$ MeV.
Note that the potential makes both of 
$0g_{9/2}$ and $2s_{1/2}$ orbits unbound.
The HO parameter $\nu$ is determined so as to maximize the average of 
the squared overlaps, $\sum_{nljm \in{F}}\langle \psi_{nljm}|\psi^{\nu}_{nljm}\rangle^2/40$
(see Ref.~\cite{Horiuchi07}). The maximum average value is 0.988 for 
set A and the resulting $\nu$ value is 0.212 fm$^{-2}$, while in set B case 
they are 0.989 and 0.203 fm$^{-2}$. Approximating the forbidden states with 
the HO wave functions is quite reasonable. 

There is no information about $^{61}$Ca. Even its stability is unknown. 
The valence neutron orbit of $^{61}$Ca belongs to
the $2n+l=4$ shell comprising the $0g$, $1d$, and $2s$ orbits.
The order of these sp orbits is under debate~\cite{Meng02, Erler12,Forssen13, Hove18, Neufcourt19, Coraggio20,Stroberg21}.
The standard shell-model filling
with the spin-orbit interaction arranges the sp levels in the order of 
$0g_{9/2}$, $2s_{1/2}$, and $1d_{5/2}$ at $A\approx 60$~\cite{BM},
while Ref.~\cite{Hagen12} predicted the inverted order of 
$2s_{1/2}$, $1d_{5/2}$, and $0g_{9/2}$. 
In the latter case, the halo structure would appear 
in the ground state of $^{62}$Ca~\cite{Hagen13}.
Following this inverted situation, we determine $V_0$ in 
Eq.~(\ref{n-60Ca.pot}) so as to set the $2s_{1/2}$ sp energy 
to be $-0.01$ MeV,
resulting in $V_0=44.03$ MeV (set A) and 41.89 MeV (set B), respectively. 
The value turns out to be slightly stronger than the standard value 
$V_0\approx 40$ MeV~\cite{BM}.

\begin{figure}[ht]
\begin{center}
\epsfig{file=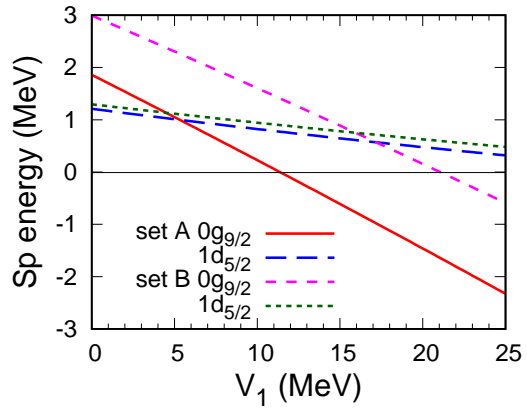,scale=1.3}
\caption{Single-particle energies of $0g_{9/2}$ and $1d_{5/2}$
  orbits of $^{61}$Ca as a function of the spin-orbit strength $V_1$.
  The diffuseness parameter, $a_{\rm WS}$, is 0.67 fm for set A and 
  0.80 fm for set B, respectively. 
  The $2s_{1/2}$ sp energy, independent of $V_1$,
  is set to $-0.01$ MeV, as indicated by the thin horizontal line.}
\label{spe61Ca.fig}
\end{center}
\end{figure}

We vary the spin-orbit strength $V_1$ in Eq.~(\ref{n-60Ca.pot}) to 
simulate different conditions for the $^{61}$Ca structure.
Figure~\ref{spe61Ca.fig} plots the sp energies of 
 the $0g_{9/2}$ and $1d_{5/2}$ orbits.
Note that the standard value of $V_1$ for $^{60}$Ca
is $ 0.44 V_0\approx 19$ MeV for set A~\cite{BM},
resulting in the sp levels of $0g_{9/2}$, $2s_{1/2}$, and $1d_{5/2}$ order.
In Fig.~\ref{spe61Ca.fig}, both sets exhibit similar sp energy 
dependence as a function of $V_1$,
though the value of $V_1$ for set B tends to be stronger
than that for set A to make $0g_{9/2}$ and $2s_{1/2}$ states degenerate due
to more diffused nuclear surface.
To fix the range of $V_1$, the following two extreme cases are considered:

{\underline {\it Vanishing spin-orbit limit.}} In Ref.~\cite{Hagen12},
the $1d_{5/2}$ and $0g_{9/2}$ sp energies of $^{61}$Ca are predicted to be 1.14 and 2.29 MeV.
To realize this situation, we need to take a very small $V_1$ value,  
$V_1\approx 0$ for set A and $\approx 5$ MeV for set B, respectively. 
We call this choice of $V_1$ the vanishing spin-orbit (so) limit.

{\underline {\it Degenerate $sg$ limit.}} 
The sp energies of the $2s_{1/2}$ and $0g_{9/2}$ orbits
are degenerate at $V_1=11.40$ MeV for set A and at $V_1=21.09$ MeV for set B. 
This choice of $V_1$ is called the degenerate $sg$ limit.
Despite both sp energies being the same, their respective 
root-mean-square (rms) radii are quite different:  36.0 and 5.04 fm for 
the $2s_{1/2}$ and $0g_{9/2}$ orbits of set A, and 36.1 and 5.28 fm for 
these orbits of set B. In both sets, 
the halo features are noticed in the $s$ states, while the rms radii 
shrink significantly for the $g$ states due to the $l=4$ centrifugal barrier. 

We examine the energy spectrum of $^{62}$Ca by varying $V_1$ between the 
vanishing and degenerate limits. 
As the outcomes of both sets A and B are qualitatively the same, we 
discuss the results obtained with set A, unless otherwise mentioned.

\subsection{Tests of FFLPG expansion}
\label{tests.sec}

Before discussing the structure of $^{62}$Ca, we evaluate the power of the 
FFLPG approach. We use the Minnesota (MN) potential~\cite{MN} for $v_{12}$ of 
Eq.~(\ref{hamiltonian}).
The MN potential, a soft-core central potential, is designed to reasonably
well reproduce the energies and sizes of $s$-shell nuclei~\cite{Suzuki08}. Since the two 
neutrons should be antisymmetric in the spin-orbital space, it is expected 
that they gain the attraction mostly in the relative $s$-wave
 and spin-singlet state. The spin-orbit and tensor components 
of $v_{12}$ are therefore expected to play an insignificant role in the 
core+$n$+$n$ model. Refer to Refs.~\cite{Suzuki04, Horiuchi06a, Horiuchi07b}
for an example of demonstrating
that the MN potential gives similar results
as the realistic $n$-$n$ ones.

Let $L$ denote the relative orbital angular momentum between
the two neutrons. The MN potential in the spin-singlet
and even $L$ channel reads 
$200 e^{-1.487r^2}-91.85e^{-0.405r^2}$ in MeV,
where $r$ is the two-nucleon distance in fm.
For the spin-triplet and odd $L$ channel it is given by 
$(200e^{-1.487r^2}-178e^{-0.639r^2})(u-1)$, where $u$ is a parameter and usually 
taken to be around 1. In what follows, $u$ is set to 1 and no interaction 
acts between the two neutrons in the spin-triplet and odd $L$ channel. 

\begin{figure}[ht]
\begin{center}
\epsfig{file=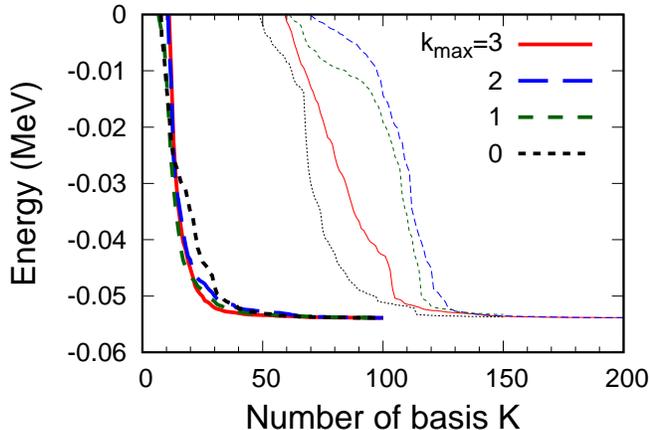,scale=1.4}
\caption{Energy convergence of the $^{60}{\rm Ca}+n+n$ system
  with $J^\pi=0^+$ as a function
  of the number of basis $K$. 
  The calculations are performed only with $l_1=l_2=0$ channel.
  Thick lines are the results with FFLPG functions, while thin lines are 
those with the pseudo-potential method.
See text for details.}
\label{conv.fig}
\end{center}
\end{figure}

To demonstrate the power of the FFLPG approach,
we focus on such a state that is 
dominated by the weakly bound $2s_{1/2}$ orbit because the approach is 
expected to have the advantage in describing nodal orbits properly.
We use the set A potential with $V_1=0$ and include the LPG bases restricted to $l_1=l_2=0$.  
Figure~\ref{conv.fig} plots the energy as a function 
of $K$, for different choices
 of $k_{\rm max}$. The candidates for the 
basis states are generated randomly 
in the interval $[0.1, 40]$ fm for $b=1/\sqrt{a}$ and in the interval 
$[0,k_{\rm max}]$ for $k$, and 
the best one is selected by the SVM algorithm.
The results with the pseudo-potential method are also shown for comparison. 
The FFLPG approach significantly improves 
the energy convergence. We confirm that $k_{\rm max}=4$ truncation 
virtually gives the same curve as the one with $k_{\rm max}=3$. 
The FFLPG calculation leads to convergence within a few tens of 
basis functions, while the pseudo-potential method needs more than 
a hundred bases to reach convergence. 
As for the pseudo-potential method,
it appears that the convergence is fastest when 
the ordinary Gaussians  with $k_{\rm max}=0$ are used, and 
there is no significant advantage in 
using the LPG bases with $k_{\rm max}>0$. 
Probably this occurs because the $k=0$ bases have the largest overlap with
the $0s_{1/2}$ and $1s_{1/2}$ forbidden states, 
with the orthogonality requirement being most efficiently met 
by a superposition of those ordinary Gaussians. 
However, the energy obtained with $K=150$  is  
$-0.0537$ for $k_{\rm max}=0$ and $-0.0533$ MeV for $k_{\rm max}=1$, respectively, 
which still misses the FFLPG energy of $-0.0539$ MeV. 
What is more serious in the pseudo-potential method is its numerical instability. 
Because the $k=0,1$ bases have in general large overlap with the forbidden states,
the calculation becomes numerically unstable due to 
the large prefactor $\lambda$ in Eq.~(\ref{proj.eq}). It is very hard to extend 
the basis size without breaking the linear 
independence of the selected bases.
With $k_{\rm max}=2,3$ this instability problem is recovered and 
we get the energy of $-0.0539$ MeV by the pseudo-potential method with $K=200$.

\begin{figure}[ht]
\begin{center}
  \epsfig{file=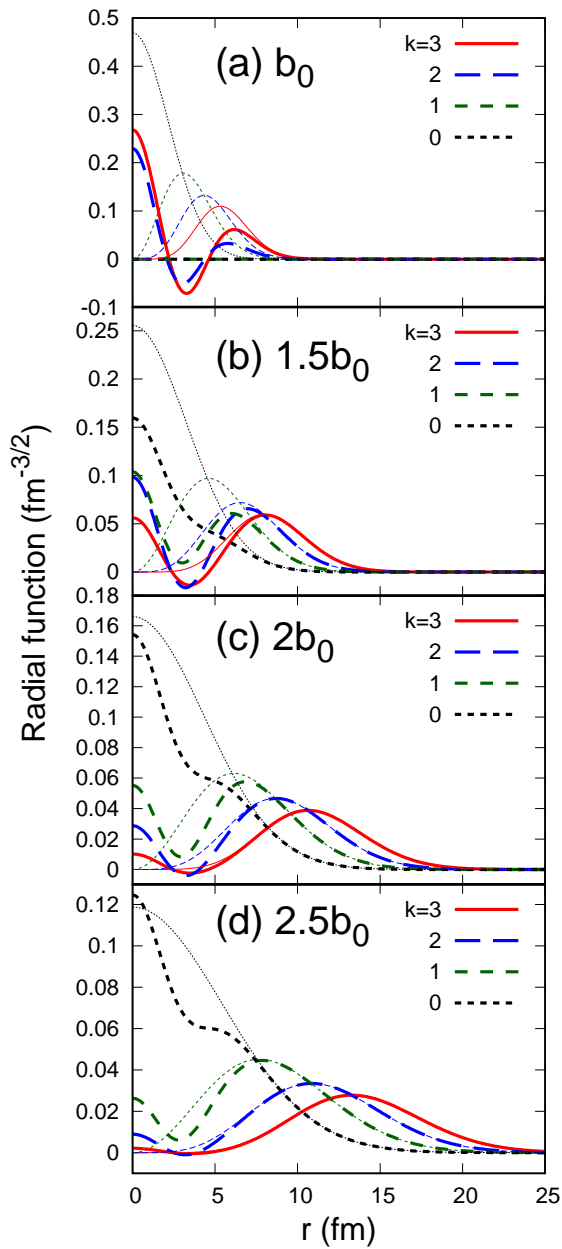,scale=1.2}  
\caption{$s$-wave radial functions of FFLPG ${\bar \phi}^a_{k0}$  
(thick curves) and LPG $\phi^a_{k0}$ (thin curves) for different $k$ 
values. The width parameter $b=1/\sqrt{a}$ is chosen to be (a) $b_0$, (b) $1.5\,b_0$, (c) $2\,b_0$, and 
(d) $2.5\,b_0$, where $b_0=1/\sqrt{\nu}=2.17$ fm with $\nu$ being
the HO parameter of the forbidden states for $^{60}{\rm Ca}+n$.
}  
\label{wfs.fig}
\end{center}
\end{figure}

Here, we should make
a comment on the role of $k$ in the LPG basis. As shown above, 
the FFLPG basis with $k_{\rm max}>0$ accelerates 
the energy convergence compared to the ordinary
Gaussians with $k=0$. To understand the reason, 
we plot in Fig.~\ref{wfs.fig} the $s$-wave radial functions with different 
$k$ values for some choices of $a$.
The thick curves are for  
$\bar{\phi}^a_{k0}$ in Eq.~(\ref{FFLPG.eq}), 
while the thin curves are for $\phi^{a}_{k0}$ 
that in general contains forbidden states. Note that the forbidden states 
are $\psi^{\nu}_{00}$ and $\psi^{\nu}_{10}$. 
In the case of $b=1/\sqrt{\nu}\equiv b_0$
the FFLPG function  thus vanishes for 
$k=0$ and 1, whereas it  
has two nodes at short distances for $k=2$ and 3 because it is orthogonal to 
the forbidden states. 
With increasing $b$, the amplitude of the FFLPG function 
becomes smaller in the inner region and finally has no node for $k=3$ at 
$b=2.5b_0$ due to small overlap with the forbidden states. 
The FFLPG basis with $k=0$ has small amplitude beyond the nuclear
surface. With increasing $k$, however, it has large amplitude beyond 
the nuclear surface and damped amplitude at short distances. 
This property meets the requirement needed for 
the neutron orbits around the core nucleus and offers the possibility 
of efficiently describing neutron orbits with large radial extension. 
The FFLPG basis is also advantageous
to gain the energy from the two-neutron interaction
because it can have large relative $s$-wave components
around the nuclear surface. 

\subsection{Application to $^{62}$Ca}
\label{62Ca.sec}

The test example presented in the previous subsection confirms that
the FFLPG offer much faster and
numerically stabler results than the pseudo-potential method.
In this subsection, we study the structure of $^{62}$Ca in 
the $^{60}{\rm Ca}+n+n$ model using the FFLPG expansion.
We take $k_{\rm max}=3$ and $l_{\rm max}=10$ 
and generate the Gaussian fall-off parameter
$b=1/\sqrt{a}$ in the interval [0.1, 40] fm.
All possible different combinations of $(l_1, j_1)$ and $(l_2, j_2)$
  are taken into account. For example, the number of combinations
  are respectively 21, 55, and 69 for $J^\pi=0^+$, $2^+$, and $4^+$ states.

\begin{figure}[ht]
\begin{center}
  \epsfig{file=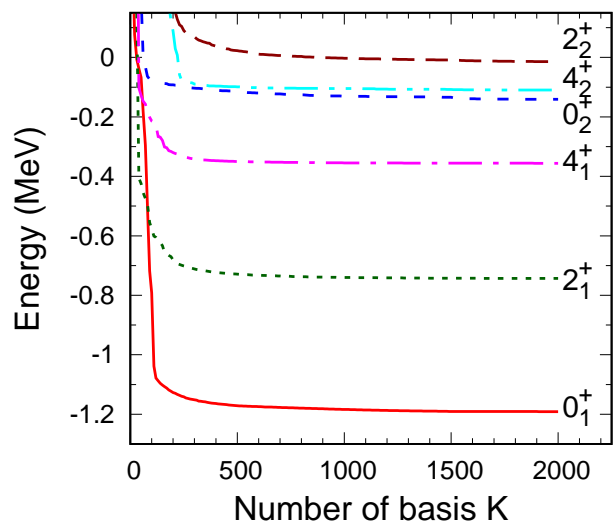,scale=1.4}
  \caption{Energies of $J^\pi=0^+, 2^+$, and $4^+$
    states of $^{62}$Ca  as a function
    of the number of basis $K$.
    The energy drawn is from the two-neutron threshold, 
    and the spin-orbit strength is $V_1=11.40$ MeV,
    set A of the degenerate $sg$ limit. }
\label{energy62Ca.fig}
\end{center}
\end{figure}

We exhibit in Fig.~\ref{energy62Ca.fig} 
the energies of the states with $J^\pi=0^+, 2^+$, and $4^+$ as a function of $K$. 
The spin-orbit strength is $V_1=11.40$ MeV, set A of the degenerate $sg$ limit. 
Note that the energy converges rapidly on a few hundred bases
for all the states. This is because all the bases are made free from 
the forbidden states.  
The fourth digit of the energy does not change on $K=1500$
for these lowest states.
The second bound states are found
for the $0^+$, $2^+$, and $4^+$ states.
We further increase the number of basis to lower
the second bound states and confirm
that they converge very well at $K=2000$.

\begin{table}
  \caption{$l_{\rm max}$-dependence of the properties of two $0^+$ states. $E$ is the energy from the $^{60}$Ca+$n$+$n$ threshold,
    $\epsilon_{2n}=\langle \sum_{i=1}^2 (T_i+U_i) \rangle$ is the sum of the two-neutron sp energies, and 
    $r_{12}=\sqrt{ \langle (\bm r_1-\bm r_2)^2 \rangle }$ is the rms relative distance between the two neutrons.
    Energy is in units of MeV and length is in units of fm. }
\begin{center}
  \begin{tabular}{cccccccc}
\hline\hline
      &&$l_{\rm max}$ & 4 & 6 & 8 & 10\\
\hline
$0_1^+$ && $E$&$-0.99$&$-1.13$&$-1.16$&$-1.19$\\
        &&$\epsilon_{2n}$&0.14 &0.33&0.38&0.43\\
        && $\left<v_{12}\right>$&$-1.13$&$-1.39$&$-1.47$&$-1.55$\\
        && $r_{12}$&7.10&6.85&6.84&6.84\\
\hline
$0_2^+$&& $E$&$-0.10$&$-0.12$&$-0.14$&$-0.14$\\
       &&$\epsilon_{2n}$ &0.17 &0.27&0.33&0.35\\
       && $\left<v_{12}\right>$&$-0.27$&$-0.39$&$-0.45$&$-0.48$\\
       && $r_{12}$&19.6&18.0&17.3&17.1\\    
\hline\hline    
\end{tabular}
  \label{lmax.tab}
  \end{center}
\end{table}

The truncation of $l_{\rm max}=10$ appears considerably large. One might question 
why so large a value is needed. Most of the sp levels are unbound in the present case,
that is, they are nonresonant continuum states. The large value 
of $l_{\rm max}$ therefore 
suggests the need to account of the continuum effect to bind the two neutrons.
Table~\ref{lmax.tab} lists the $l_{\rm max}$ dependence of the basic 
properties of the $0_1^+$ and $0_2^+$ states.
As the table indicates clearly, 
$\epsilon_{2n}$ increases, while $\langle v_{12}  \rangle$ gets more attractive 
as a function of $l_{\rm max}$. The $l_{\rm max}$ dependence of the two-neutron rms distance, 
$r_{12}$, shows an apparent correlation with $\langle v_{12}  \rangle$. 
Although we confirm that the truncation with $l_{\rm max}=10$ takes into account 
most of the continuum effect, it appears that there may be still some room to  
improve the convergence by increasing $l_{\rm max}$ further. 

In addition to the six states drawn
in Fig.~\ref{energy62Ca.fig}, we obtain two bound states
with $6^+$ and $8^+$. Table~\ref{62Ca.tab} summarizes the energies, the rms neutron radii, and 
the occupation probabilities for those bound states. Note that 
the $0g_{9/2}$ and $2s_{1/2}$ sp states are set to be degenerate.
The ground state exhibits nonhalo structure, occupying the $(g_{9/2})^2$ configuration 
by 94\%. The configuration also produces
the $2^+_1, 4^+_1, 6^+$, and $8^+$ bound states
that have almost the same structure as the ground state. Halo 
structure is realized as the $0^+_2$ and $4^+_2$ states. Both states have the 
rms neutron radius, $r_{2n}$, larger than 10 fm. As shown by the occupation 
probability, the $0^+_2$ state may be called an $s$-wave two-neutron halo. 
It should be noted, however, that its $r_{2n}$ value of about 13 fm is by far 
smaller than the rms radius of the $2s_{1/2}$ sp orbit, which is about 36 fm. 
This dramatic reduction is of course due to the correlated motion of 
the two neutrons. The one-neutron halo character
of the $4^+_2$ state is brought about 
by the coupling of the $g_{9/2}$ neutron
with the $s_{1/2}$ neutron, as revealed 
by the occupation probabilities.
This suggests that both states of
  two- and one-neutron halo structure can coexist 
  within a narrow energy spacing.
Despite the fact that it  
is barely bound, the $2^+_2$ state shows no characteristics of halo structure. 
Its $r_{2n}$ value is considerably large, 6.77 fm, but it is by far smaller 
than those of the $0^+_2$ and $4^+_2$ states.
Note that the recoil kinetic energy is not negligible
as the binding energy is small. This is because the main configuration
of the $2^+_2$ state is $(g_{9/2}d_{5/2})$.

Table~\ref{62Ca.tab} also lists the decomposition of the energy $E$ into the sp energy,
the recoil kinetic energy, the two-neutron interaction energy, and the rms 
two-neutron distance. The positive value of 
$\epsilon_{2n}$ indicates that the two neutrons move mostly in the continuum 
states. The recoil kinetic energies are small due to the factor $1/60m_N$.  
Clearly, the two-neutron interaction $v_{12}$ plays a decisive role to 
make the two neutrons bound.
At a closer look, the $2_2^+$ state shows
  the largest sp energy and a relatively large energy
  gain from the two-neutron interaction energy.
  The state is realized by the coupling of the continuum states
  such as the $d_{5/2}$ and other high $l$ orbits
  due to the two-neutron correlation.
  
The $(g_{9/2})^2$ dominance in the ground state
can be explained by the pairing antihalo effect pointed out
in Ref.~\cite{Masui20}. A combination of the higher angular momentum
states is more advantageous to gain energy from the pairing.
In fact, $\left<v_{12}\right>$ values 
are $-1.55$ and $-0.48$ MeV for $0^+_1$ and $0_2^+$, respectively.
To make the ground state  a halo, 
the energy difference between the $2s_{1/2}$ and $0g_{9/2}$ orbits
should be sufficiently larger than the energy gain
of the two-neutron interaction with the $(g_{9/2})^2$ configuration.

As mentioned above, the $4^+_2$ state is found to have one-neutron halo
structure constructed dominantly from the $(s_{1/2}g_{9/2})$ configuration.
That configuration would suggest a doublet state with $5^+$.
However, its dominant configuration is in the spin-triplet and odd $L$
channel. Possible existence of this unnatural-parity state
crucially depends on the choice of $u$ parameter.
Since we set $u=1$, $v_{12}$ vanishes and no doublet state appears.
To be more definitive about its existence, we have to test
other realistic $n$-$n$ potentials.

\begin{table*}
  \caption{
Properties of the spectrum of $^{62}$Ca. The sp energies of $2s_{1/2}$ and $0g_{9/2}$ orbits are set to be $-10$ keV using the 
parameters of the degenerate $sg$ limit of set A.
The sp energy of the $1d_{5/2}$ orbit is 0.77 MeV.
See Table~\ref{lmax.tab} for the definitions of $E, \ \epsilon_{2n}$, and $r_{12}$. $r_{2n}=\sqrt{\langle \frac{1}{2} (r_1^2+r_2^2)\rangle}$ is the rms neutron 
radius, and $\langle T_{\rm rec}\rangle=\langle \frac{1}{60m_N}\bm p_1\cdot \bm p_2 \rangle$ is the recoil kinetic energy.  $P_{xy}$ denotes the occupation probability of finding $x, y$ sp orbits, where $x$ and $y$ stand for $s_{1/2},\ g_{9/2}, \ d_{5/2}$, respectively, and $\Delta P=1-\sum_{xy}P_{xy}$.}
\begin{center}
\begin{tabular}{ccccccccccccccc}
\hline\hline
$J^\pi$ & $E$(MeV)&$r_{2n}$(fm)&
$P_{ss}$&$P_{gg}$&$P_{dd}$&$P_{sg}$&$P_{sd}$&$P_{gd}$&$\Delta P$&
$\epsilon_{2n}$(MeV)& $\left<T_{\rm rec}\right>$(MeV)& $\left<v_{12}\right>$(MeV)&$r_{12}$(fm)\\
\hline
$0_1^+$&$-1.19$&5.08&0.01&0.94&0.02&--&--&--  &0.02& 0.43 & $-$0.07 & $-$1.55  &6.84\\
$2_1^+$&$-0.74$&5.12&--&0.86&0.01&--&0.01&0.09&0.03&  0.40 & $-$0.06 & $-$1.08 &6.96\\
$4_1^+$&$-0.36$&5.35&--&0.87&0.00&0.09&--&0.03&0.01&  0.14 & $-$0.03 & $-$0.47 &7.43\\
$6^+$&$-0.22$&5.03&--&0.99&--&--&--&0.01      &0.01&  0.04 & $-$0.02 & $-$0.24&7.05\\
$8^+$&$-0.21$&5.02&--&0.99&--&--&--&--        &0.01& 0.06 & $-$0.01 & $-$0.26 &7.06 \\
$0_2^+$&$-0.14$&12.8&0.91&0.02&0.05&--&--&--  &0.03& 0.35 & $-$0.01 & $-$0.48 &17.1\\
$4_2^+$&$-0.11$&10.1&--&0.10&0.00&0.87&--&0.02&0.01 & 0.08 & $-$0.01 & $-$0.18& 14.3\\
$2_2^+$&$-$0.014&6.77&--&0.13&0.03&--&0.14&0.64&0.06&  1.04& $-$0.04 & $-$1.02&9.09  \\
\hline\hline
\end{tabular}
\label{62Ca.tab}
  \end{center}
\end{table*}
    
As discussed above, when the spin-orbit strength $V_1$
of the $n$-$^{60}$Ca potential is taken to be the degenerate $sg$ limit,
the $J^{\pi}=0^+$ halo structure appears as the 
excited state. It is interesting to 
examine how the $0^+$ state changes if the spin-orbit strength is weakened  
towards the vanishing so limit. We introduce a multiplicative 
factor $f$ and set the spin-orbit strength
as $V_1=f V_1^\prime \, (0\leq f \leq 1)$ with $V_1^\prime=11.40$ MeV for set A.
The limit of $f\approx 0$ is the case used in Ref.~\cite{Hagen12}, where 
the energy difference between the $2s_{1/2}$ and $0g_{9/2}$ orbits is large
(see Fig.~\ref{spe61Ca.fig}) 
and only one bound state of two-neutron halo structure is predicted. 

\begin{figure}[ht]
\begin{center}
  \epsfig{file=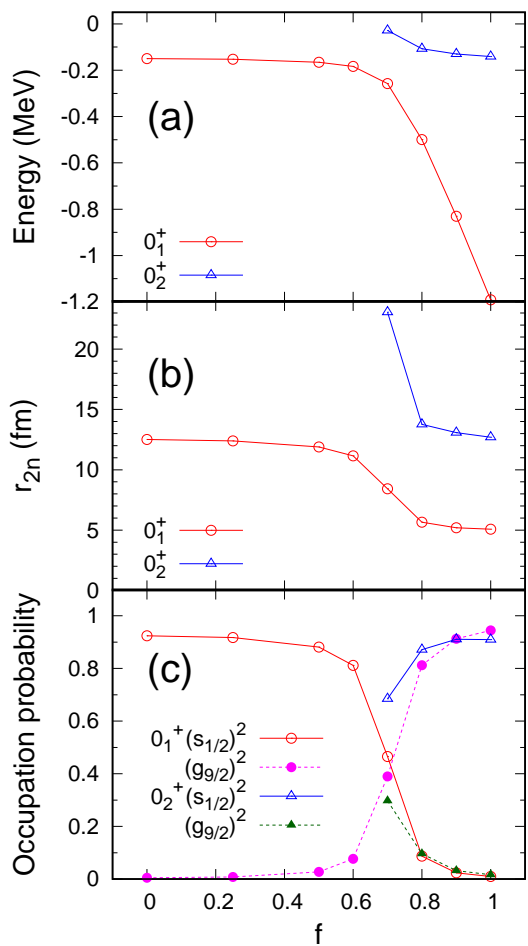,scale=1.2}
  \caption{(a) Energies, (b) rms two-neutron radii,
    and (c) occupation probabilities of the $0_1^+$ and $0_2^+$ states
    of $^{62}$Ca as a function of $f$. The spin-orbit strength 
  is taken to be $V_1=fV_1^\prime$ with $V_1^\prime=11.40$ MeV for set A.}
\label{62Ca.fig}
\end{center}
\end{figure}

Figure~\ref{62Ca.fig} (a) displays the energy of the $0^+$ state 
as a function of $f$.  Only one $0^+$ state appears for $f <0.7$ and the 
existence of the second $0^+$ state is possible for $f\geq 0.7$.
Figure~\ref{62Ca.fig} (b) shows the corresponding rms neutron radii.
The halo structure having a radius larger than 10 fm
emerges in the ground state for $f<0.7$. For  $f>0.7$ 
the $0_2^+$ state exhibits the two-neutron halo structure, whereas 
the ground state turns out to be a compact state.
At $f=0.7$ the ground state shows intermediate structure between 
halo and compact states.
The rms radius of the $0_2^+$ state is extremely large due to the small binding energy of $-$0.03 MeV.
This behavior can be understood by showing the occupation probabilities.
Figure~\ref{62Ca.fig} (c) shows 
the occupation probabilities of finding $(s_{1/2})^2$ and $(g_{9/2})^2$ 
components in the $0^+$ states. 
The contributions of $(d_{5/2})^2$ component 
for the $0^+_1$ state are at most 0.09 at $f=0.7$  and 
other contributions are less than 0.05 in total.
For the $0_2^+$ state, those contributions are less about 0.01.
 As expected, the $(s_{1/2})^2$ component dominates in
the ground state at $f=0$, resulting in  a large rms radius of 12 fm,
almost the same structure as the $0_2^+$ state at $f=1$.
By increasing $f$, i.e., reducing the energy difference
between the $2s_{1/2}$ and $0g_{9/2}$ orbits,
the occupation probability of the $0g_{9/2}$ orbit increases gradually
and rises suddenly for $f>0.6$.
The rms radius decreases simultaneously with the growing occupation of
the $0g_{9/2}$ orbit that has a much smaller radius.
An almost equal mixing
of the $(s_{1/2})^2$ and $(g_{9/2})^2$ configurations
occurs at $f=0.7$.
Here the energy difference between the $2s_{1/2}$ and $0g_{9/2}$ orbit
is approximately 0.5 MeV, which is 
comparable to the difference
of $\left<v_{12}\right>/2$ for
the two $0^+$ states as shown in Table~\ref{62Ca.tab}.
Finally, the ground state becomes $g_{9/2}$-dominant at $f=1$, where 
the sp energies of the $2s_{1/2}$ and $0g_{9/2}$ orbits are degenerate,
while the $0_2^+$ state exhibits two-neutron halo structure 
consisting of the $(s_{1/2})^2$ configuration.

\subsection{Probable halo ground state of $^{72}$Ca}
\label{72Ca.sec}

An extension to the $^{70}{\rm Ca}+n+n$ model is straightforward.
The reader is referred to Ref.~\cite{Hove18} for the structure of $^{72}$Ca
by the hyperspherical method. 
We follow the case of $^{62}$Ca starting from discussing some constraints 
on the phenomenological WS potential parameters.

The $n$-$^{70}{\rm Ca}$ potential should bind 
the $0g_{9/2}$ orbit due to the assumption of the $N=50$ core. 
However, no bound $0g_{9/2}$ orbit is generated if we assume 
the parametrization of Ref.~\cite{BM}. Instead of searching for such a 
potential that binds the $0g_{9/2}$ orbit, we simply assume that 
all the occupied neutron orbits including the $0g_{9/2}$ orbit are described 
by the HO functions with the oscillator parameter $\nu'$ that is scaled 
from the $^{60}$Ca parameter $\nu$ by $\nu^\prime=(60/70)^{1/3}\nu$. 
The sp energy spectrum of $^{71}$Ca 
with respect to $V_1$ is virtually the same
as Fig.~\ref{spe61Ca.fig} except for the absence of the $0g_{9/2}$ level.

To bind the $2s_{1/2}$ orbit at $-0.01$ MeV, the strength $V_0$ turns out 
to be 40.26 MeV for set A and 38.32 MeV for set B.
Since $V_0$ gets weaker than the $^{60}$Ca-$n$ case, 
we have only one bound $0^+$ state that has two-neutron halo
structure if $V_1$ is taken to be the same as 
that of $^{61}$Ca: The energy, rms neutron radius, and $(s_{1/2})^2$ and 
$(d_{5/2})^2$ occupation probabilities are $-0.14$ MeV, 12.7 fm, and 
0.89 and 0.08 for set A, whereas for set B they are $-0.13$ MeV, 13.3 fm, 
and 0.87 and 0.11, respectively. The $n$-$^{70}$Ca potential used here makes 
the energy gap between the $2s_{1/2}$ and $1d_{5/2}$ orbits too large to 
mix those sp states.
To make the $1d_{5/2}$ orbit bound  is very 
unlikely because $V_1$ has to be taken more than two times larger 
than the standard value.
Within the present phenomenological $n$-$^{70}{\rm Ca}$ potential,
the halo structure emerges
if the energy of the $2s_{1/2}$ state is close to zero,
which is the same conclusion drawn in Ref.~\cite{Hove18}.

\section{Conclusions and prospects}
\label{conclusions.sec}

We have developed a forbidden-state-free locally peaked Gaussian
expansion method to describe the weakly bound correlated neutron motion
around a heavy core. The power of this approach has been tested
by taking examples of
the weakly bound three-body systems $^{60,70}{\rm Ca}+n+n$.
The expansion proposed here accelerates 
the energy convergence much faster 
and offers numerically by far stabler results 
than the pseudo potential projection method~\cite{Kukulin78}.
The present method allows us to predict very weakly bound excited states.
The energy spectra of unknown $^{62,72}$Ca have been calculated
to discuss the possibility of  halo-structure emergence by varying
the single-particle levels of unknown $^{60,70}{\rm Ca}+n$ systems.

In the extreme single-particle level ordering of $2s_{1/2}, 1d_{5/2}$, and 
$0g_{9/2}$ orbits predicted by Ref.~\cite{Hagen12},
only one bound state is found in $^{62}$Ca,
exhibiting $s$-wave two-neutron halo structure. The emergence of the halo 
structure in the ground state in fact strongly depends on the energy 
difference between the $2s_{1/2}$ and $0g_{9/2}$ orbits for $^{62}$Ca and 
between the $2s_{1/2}$ and $1d_{5/2}$ orbits for $^{72}$Ca. 
Under the very limited condition that the $2s_{1/2}$ and $0g_{9/2}$ orbits are 
bound and degenerate, however, the ground state becomes a $0^+$ nonhalo state 
dominated by the $0g_{9/2}$ orbit, consistent with the pairing 
antihalo effect~\cite{Masui20}.
In that case, a $0^+$ two-neutron halo state appears 
slightly above the ground state. In addition a $4^+$ one-neutron halo state 
appears at almost the same excitation energy as the $0^+$ excited state. 
Apparently experimental information on the single-neutron
levels of $^{60,70}{\rm Ca}+n$ system
is crucially important to identify the structure
of these neutron-rich Ca isotopes. 

The method proposed here can straightforwardly be extended 
to core plus more-nucleon systems.
Since it is advantageous to describe localized orbits near
the surface of the core, it is interesting to apply
the method for alpha-decay phenomena of heavy nuclei.
The result of Ref.~\cite{Varga94} for the $^{212}$Po problem
could be improved with locally peaked Gaussians.
A systematic analysis of the degree of the alpha clustering
near the nuclear surface is interesting as
it has recently been realized in alpha-knockout reactions
on Sn isotopes~\cite{Tanaka21}.

\acknowledgements

This work was in part supported by JSPS KAKENHI Grant No. JP18K03635
and the Collaborative Research Program 2021,
Information Initiative Center, Hokkaido University. 
For partial support, L.T. thanks Funda\c c\~ao de Amparo \`a Pesquisa do Estado de S\~ao Paulo (FAPESP) (Projects No. 2017/05660-0 and No. 2015/11828-5)
and Conselho Nacional de Desenvolvimento Cient\'\i fico e 
Tecnol\'ogico [Projects No. 304469/2019-0 and No. 464898/2014-5 (INCT-FNA)];
M.A.S. thanks 
Coordena\c c\~ao de Aperfei\c coamento de Pessoal de N\'\i vel Superior.
Y.S. is indebted to L.T. for his generous invitation
to S\~ao Paulo for September to December 2015 through the FAPESP grant,
which made it possible to start this collaboration.

\end{document}